\documentclass[journal=jpcafk,manuscript=article]{achemso}
\usepackage[version=3]{mhchem}
\usepackage{titlecaps}
\Addlcwords{the of into}
\usepackage{chemformula} 
\usepackage[T1]{fontenc} 
\usepackage{enumitem}
\usepackage{graphicx}
\usepackage{titlecaps}
\Addlcwords{a an the of on at with for and in into from by without}

\author{Markus
  Meuwly}\email{m.meuwly@unibas.ch}\affiliation[University of Basel]
       {Department of Chemistry, University of Basel,
         Klingelbergstrasse 80, 4056 Basel, Switzerland}

\date{\today}

\title[]{Atomistic Simulations for Reactions and Spectroscopy in the
  Era of Machine Learning - {\it Quo Vadis?}}

\abbreviations{}
\keywords{}

\begin{document}

\date{\today}

\begin{abstract}
Atomistic simulations using accurate energy functions can provide
molecular-level insight into functional motions of molecules in the
gas- and in the condensed phase. Together with recently developed and
currently pursued efforts in integrating and combining this with
machine learning techniques provides a unique opportunity to bring
such dynamics simulations closer to reality. This perspective
delineates the present status of the field from efforts of others in
the field and some of your own work and discusses open questions and
future prospects.
\end{abstract}

\section{Introduction}
``Atomistic Simulations'' provide molecular-level information into
chemical and biological phenomena. This is largely due to the
possibility to explicitly follow the dynamics of molecular systems
with the full information about observables contained in the time
series of positions and velocities. The ``translation'' between the
output from a molecular dynamics (MD) simulation (positions and
velocities as a function of time) is accomplished by virtue of
statistical mechanics, in particular the correlation function
formalism introduced in the 1960s by Zwanzig\cite{zwanzig:1965} and
Gordon.\cite{gordon:1968}\\

\noindent
The nuclear dynamics - whether described at a classical or at a
quantum level - is primarily governed by the underlying energy
function $V(\vec{x})$ where $\vec{x}$ encompasses all $3N$ coordinates
of an $N-$particle system. Most rigorous for such studies will be to
use an exact or near-exact solution of the electronic Schr\"odinger
equation together with corresponding forces for each configuration
$\vec{x}$. Doing this ``on the fly'' is typically not possible except
for systems with a small number of electrons, and not on extended time
scales. Hence, such {\it ab initio} MD (AIMD) methods often use
approximations at the density functional theory (DFT) or semiempirical
level of theory and typically within a mixed quantum
mechanics/molecular mechanics (QM/MM)
framework.\cite{cui:2016,cui:2021}\\

\noindent
Since about 10 years the field of computational and theoretical
chemistry has seen a profound change in how we think about inter- and
intra-molecular interactions. One milestone has been the realization
that with a sufficient number of samples the explicit solution of the
electronic Schr\"odinger equation can be replaced by a data-driven
approach based on machine learning (ML).\cite{rupp:2012} This approach
consistently applies principles rooted in statistical learning
theory.\cite{vapnik:1998} A recent series of reviews delineates the
present state of the art of ML-based methods for force
fields\cite{unke:2021}, for small molecules\cite{manzhos:2020}, and
for chemical reactions.\cite{MM:2021} It is of interest to mention
that for intermolecular interactions such ML-based methods have been
used for a considerably longer
time,\cite{ho96:2584,hollebeek.annrevphychem.1999.rkhs} especially for
triatomics,\cite{MM.heh2:1999} based on early work by Aronszajn on
reproducing kernel Hilbert spaces (RKHS).\cite{aronszajn1950theory}\\

\noindent
One of the particular benefits of atomistic simulations in the context
of experiments is the possibility to relate observables with
particular structural features of a system. A topical example concerns
infrared (IR) spectroscopy for gas- or condensed-phase systems. At
ambient conditions even small peptides sample different conformations
each of which exhibits potentially conformer-specific IR spectra. This
applies to both, experiments in the gas phase and in
solution.\cite{zwier:2006,rizzo:2009,amadei:2010} The question thus
arises to what conformational substate a particular set of spectral
lines belongs and whether a unique correspondence between structure and
spectroscopy can be obtained. However, assigning the spectroscopic
features to a particular conformational substate remains a
challenge. One possibility to address this problem consists in
carrying out extensive electronic structure calculations which,
however, are time consuming and usually only applicable to gas-phase
systems. Alternatively, MD simulations with improved force fields can
be used to determine the underlying structural features by comparing
computed and experimentally measured IR
spectra.\cite{Tokmakoff2018,MM.ala3:2021}\\

\noindent
Similarly, a beneficial interplay between experiment and computation
has clarified the relationship between structure and spectroscopy for
protonated water clusters.\cite{jordan:2005,chang:2005,zeng:2021}
Often, such experiments are carried out in a supersonic jet combined
with mass selection of particular cluster sizes which generates an
ensemble of structures that is probed by light. As there is a
distribution of cluster geometries that is frozen in the expansion,
the spectroscopic signatures are those of a number of - typically -
low-energy isomers and not of one single conformer. What spectral
features belong to one particular conformation can be clarified by
using hole-burning experiments\cite{zwier:2008} and conformer-specific
spectra are then determined from electronic structure
calculations. Such an approach has proven very powerful, in particular
for small protonated water
clusters.\cite{bowman:2017,bowman:2017.jacs} Very recently, the
infrared spectroscopy of H$^+$(H$_2$O)$_{21}$ was assigned to
particular structural features using Second-order Vibrational
Quasi-Degenerate Perturbation Theory (VQDPT2). Based on the agreement
between experiment and simulations particular features in the
experimental spectra in the high-frequency (OH-stretch) region could
be assigned to specific bonding patterns in the protonated water
cluster.\\

\noindent
Combining computational and experimental approaches is also valuable
for chemical reactions in the gas and in the condensed phase. Typical
question arising in this context are A) whether reactions proceed
stepwise or concerted, e.g. for phosphate
transfer,\cite{aqvist:1999,klahn:2006,duarte:2015}
Menshutkin-type,\cite{Menschutkin1890,allfrey:1964,Schubert2003,Castejon1999,Castejon2001,MM.mensh:2021}
or Diels-Alder
reactions,\cite{houk95a,domingo14a,MM.da:2019,MM.da:2021} or B) for
unimolecular reactions whether I) decomposition involves one or
several competing routes, II) a potential barrier is involved, III) it
simply progresses by stretching a chemical bond until it breaks, or
IV) a combination of II) and III) is at play which was coined
``roaming''.\cite{bowman:2011} Also, competing pathways can occur and
have been found from atomistic
simulations.\cite{yanai:1997,MM.ngb:2013,glowacki:2015}\\

\noindent
In what follows an account of the status and future prospects of
atomistic simulations with a particular focus on classical MD
simulations in the context of experiments is given. Findings and
relevant advances from recent studies in our group and others in the
field are highlighted and open questions that require and spur future
developments are formulated and discussed.

\section{Vibrational Spectroscopy}
Optical spectroscopy is a versatile and powerful method to interrogate
and characterize the structural dynamics of systems in the
gas\cite{oomens:2006,wolk:2014} and in the condensed
phase.\cite{zanni:2001,diller:2002,anfinrud:2003,decatur:2006,zheng:2007,kim:2009}
It has been proposed (for a cyano-substituted ligand bound to WT and
mutant human aldose reductase)\cite{boxer:2006} and explicitly
demonstrated by simulations (for cyano-benzene bound to
lysozyme)\cite{MM.lys:2017} that IR spectroscopy can be used to infer
ligand binding in particular protein pockets. The possibility to
relate changes in the infrared spectroscopy with changes in
intermolecular binding modes and/or interaction strengths is an
exciting prospect for functional studies of biomolecules under
realistic conditions, including solvent and ions.\\

\noindent
One application of vibrational spectroscopy is positionally resolved
infrared spectroscopy to relate structural features and local/global
dynamics with spectroscopic responses. For this, suitable
spectroscopic probes are required. One such label is azidohomoalanine
(AHA)\cite{hamm:2012} for which experiments demonstrated that it can
be used for site-specific information on the recognition site between
the PDZ2 domain and its binding partner.\cite{stock:2018} AHA absorbs
at around $\sim 2100$ cm$^{-1}$ with a substantial extinction
coefficient ($\sim 400$ M$^{-1}$cm$^{-1}$).\cite{hamm:2012} Attaching
--N$_3$ to alanine (to give AlaN$_3$) or AHA and incorporation at
almost any position of a protein has been demonstrated following known
expression techniques.\cite{bertozzi:2002} Furthermore, attachment of
an --N$_3$ probe is a spatially small modification and the chemical
perturbations induced are expected to be small. This makes AlaN$_3$
and AHA worthwhile modifications to probe local protein dynamics.\\

\noindent
Exhaustive labeling of all 14 alanine residues in WT human Lysozyme
provided dynamical information in a position-resolved
manner.\cite{MM.lys:2021} These simulations were carried out with a
conventional protein force field but the energetics of the
spectroscopic reporter (N$_3^-$) was a reproducing kernel Hilbert
space\cite{ho1996general,MM.rkhs:2017} based on high-level
PNO-LCCSD(T)-F12\cite{lccsd-schwilk-2017,lccsdf12-schwilk-2017}
calculations together with the aug-cc-pVTZ basis set. This is akin to
a mixed quantum mechanics/molecular mechanics (QM/MM) procedure with
the advantage that a) evaluating the RKHS is computationally similarly
efficient as computing energies and forces for an empirical force
field which allows to carry out multi-nanosecond simulations for such
a system and b) the quality of the RKHS is that of the underlying
quantum chemical method.\\

\noindent
Using the same force field for the azide probe attached to every
alanine residue in Lysozyme leads to frequency maxima covering a range
of $\sim 15$ cm$^{-1}$.\cite{MM.lys:2021} This compares favourably
with an experimentally reported frequency span of $\sim 10$ cm$^{-1}$
for replacements of Val, Ala, or Glu by AHA in the PDZ2
domain.\cite{hamm:2012} The frequency fluctuation correlation
functions (FFCFs) for the N$_3^-$ asymmetric stretch mode decay on two
time scales. One is in the sub-picosecond regime and quite universal
at $\tau_1 \sim 0.1$ ps whereas the longer decay time ranges from
$\sim 1$ to $\sim 10$ ps depending on the position of the
label. Experimentally, correlation times of 3 ps have been measured
for AHA.\cite{hamm:2012} Another relevant experimental observable is
the inhomogeneous contribution which arises from dynamics that relaxes
on time scales longer than $\tau_2$. Interestingly, the magnitude of
the static component was found to correlate qualitatively with the
degree of hydration of the spectroscopic probe.\cite{MM.lys:2021} \\

\noindent
Another application of simulation and experimental approaches for
vibrational spectroscopy is the characterization of the conformational
ensemble sampled by peptides and proteins. A widely studied model
system in this context is trialanine
(Ala$_3$).\cite{woutersen:2000,Woutersen,Hamm.pnas.2001,Schweitzer-Stenner.jacs.2001,Woutersen:2001,Mu.jpcb.2002,Graf2007,Gorbunov2007,Oh2010,Xiao2014,Tokmakoff2018,MM.ala3:2021}
The conformational ensemble from most of the investigations is
dominated by the poly-proline II (ppII) structure, often followed by a
$\beta-$sheet conformation and more rarely some right-handed
$\alpha-$helical structure. In a notable combined experimental and
simulation study the infrared spectroscopy in the amide-I region was
used to refine the conformational ensemble. This was done by Bayesian
ensemble refinement of the conformational ensemble such as to best
reproduce the experimentally measure IR spectra.\cite{Tokmakoff2018}
The refinement effectively reweights a reference distribution with
associated basin-specific IR spectra to better describe the
experimentally observed spectra. Interestingly, comparable changes in
the conformational ensemble were found when going from a conventional
point charge force field to a multipolar representation of the amide
group although the ppII structure is probably overstabilized in these
simulations.\cite{MM.ala3:2021}\\

\noindent
Such studies lay the foundation for more functionally relevant studies
of biomolecules such as the A$\beta$ amyloids or insulin. 1d- and
2d-infrared spectroscopy have been used to probe structural features
and the dynamics of different amyloids.\cite{moran:2014} One
distinguishing feature of the 1-dimensional spectra is a narrow,
intense absorption between 1615 and 1630 cm$^{-1}$ which is shifted to
the red of the typical $\beta-$sheet amide-I bands. Heavy atom
labeling of individual -CO groups allows to assign the spectroscopic
response to particular parts of the protein. Using isotope labeling
together with 2d-IR spectroscopy and computed spectra from MD
simulations allowed to distinguish between a $\beta-$arch and a
$\beta-$turn configuration of polyQ fiber
segments.\cite{buchanan:2014}\\

\noindent
Infrared spectroscopy is also potentially useful for following protein
assembly and disassembly. Insulin, which is key for the glucose cycle,
binds to the insulin receptor in its monomeric form. However, in the
body the hormone is stored as zinc-bound hexamers each of which
consists of three homodimers. For human WT insulin the stabilization
of the dimer with respect to two separated monomers has been
determined experimentally ($\Delta G = -7.2$
kcal/mol)\cite{strazza:1985} and all-atom simulations in explicit
solvent confirm these measurements ($\Delta G$ ranging from --8.4 to
--11.9 kcal/mol from free energy simulations and --12.4 kcal/mol along
the minimum energy
path).\cite{MM.insulin:2005,MM.insulin:2018,bagchi:2019} However, for
pharmaceutical applications modified insulins have been used and are
being designed for which dimerization free energies are not
available.\\

\begin{figure}
 \begin{center}
 \resizebox{0.99\columnwidth}{!}
           {\includegraphics[scale=0.1,clip,angle=0]{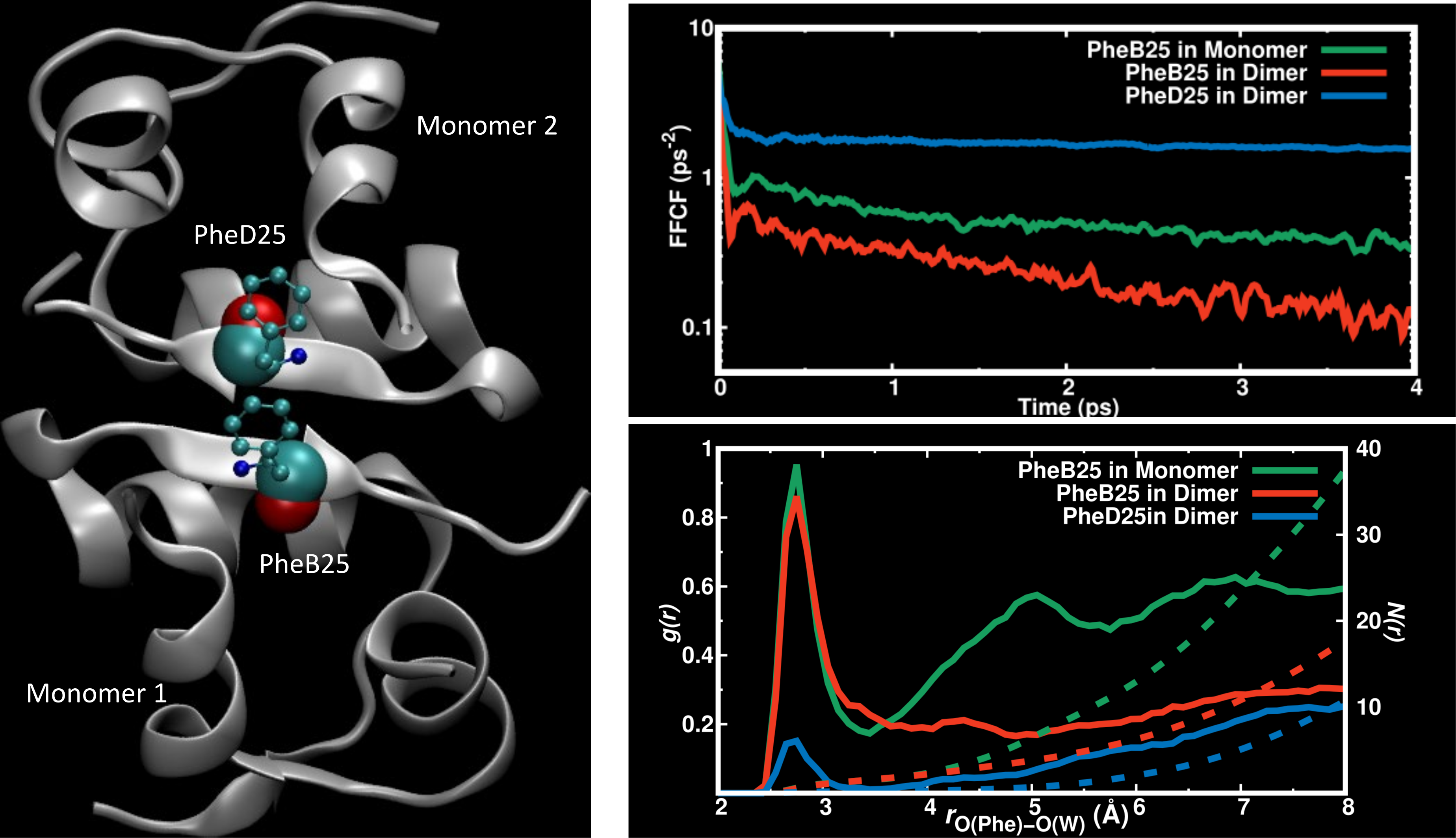}}
           \caption{Structural dynamics involving the PheB25 residue
             in monomeric and dimeric insulin. Left panel: The
             structure of the insulin dimer with residues PheB25 and
             PheD25 as highlighted by the blue spheres and the -CO
             group as van der Waals spheres. Upper right panel: the
             frequency fluctuation correlation function for PheB25 in
             the monomer (green), and for PheB25 and PheD25 in the
             dimer (red and blue). Lower right panel: the hydration of
             the -CO group for PheB25 in the monomer (green solid
             line), and for PheB25 and PheD25 in the dimer (red and
             blue solid lines) together with the total number of water
             molecules $N(r)$ as dashed lines with the same color
             code.}
\label{fig:fig1}
\end{center}
\end{figure}

\noindent
Experimental studies have provided insights about parts of the
dissociation pathway between the dimer and the separated
monomer. Two-dimensional IR studies in the amide-I region coupled with
$T-$jump for the dimer indicate that monomers within the dimer
rearrange on the 5 to 150 $\mu$s time
scale.\cite{ganim:2010,tokmakoff:2016} Between 250 and 1000 $\mu$s the
$\beta-$sheet structure at the dimer interface is lost. Complementary
to this, time-resolved X-ray scattering studies reported the
population of two intermediate dimer states, D$_1$ and D$_2$, on the
310 ns and 900 ns time scale.\cite{rimmerman:2017} Although these
studies are valuable from a spectroscopic perspective, they only
provide limited structural information.\\

\noindent
Computational infrared spectroscopy based on MD simulations provides a
means to link structural dynamics and spectroscopy. For insulin
monomer and dimer, MD simulations using validated multipolar force
fields have shown that amide-I spectroscopy at the dimerization
interface - involving residues PheB24, PheB25, and TyrB26 - depends on
the aggregation state,\cite{MM.insulin:2020} see Figure \ref{fig:fig1}
for Phe25 as an example. In other words, the IR response of the -CO
reporters for residues B24 to B26 from the isolated monomer in
solution differs from that of the two monomers in the dimer. In
addition, the position of the maxima in the IR absorption and the
dynamics of the symmetry-equivalent residues [B24,D24], [B25,D25], and
[B26,D26] is not identical. This suggests that the two monomers in the
dimer are dynamically not identical on the time scale of the
simulations. Figure \ref{fig:fig1} demonstrates that the FFCF for
PheB25 in the monomer differs from that of PheB25 and PheD25 that also
differ from one another in the dimer. These findings are also
consistent with X-ray experiments which found the two monomers to be
not exactly symmetric.\cite{BakerPhilos1988} In this context it is of
interest to note that ``..[the] ability of the insulin molecule to
adopt different conformations may be an important factor in the
expression of its biological activity..''.\cite{dodson:1979} Hence,
the dynamical asymmetry found in the present work for WT and mutant
insulin dimers may have functional implications. Overall, these
findings also support the conclusions from recent MD simulations which
report that early along the pathways between insulin dimer and two
separate monomers the two monomers behave
asymmetrically.\cite{dinner:2020}\\

\begin{figure}
 \begin{center}
 \resizebox{0.99\columnwidth}{!}
           {\includegraphics[scale=0.1,clip,angle=0]{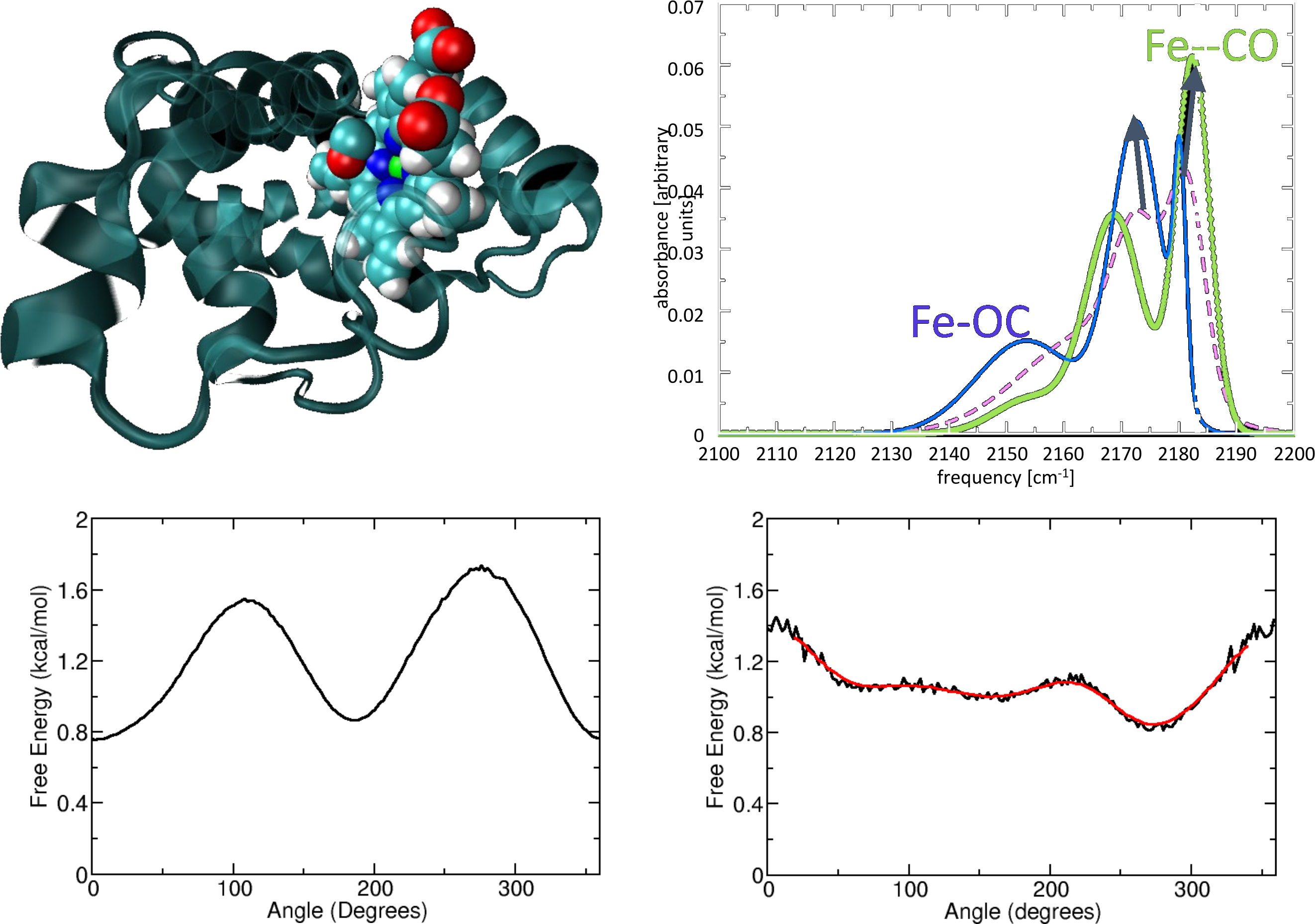}}
           \caption{The infrared spectroscopy of photodissociated CO
             in myoglobin (Mb). Upper left panel: Structure of the
             protein with heme and CO as van der Waals spheres and the
             heme-Fe as a green sphere. Upper right panel the infrared
             spectrum from a 2 ns simulation using a multipolar
             representation for the CO molecule. The dashed line is
             for the full trajectory whereas the blue and green solid
             lines are for the parts of the trajectory during which
             the Fe--OC and Fe--CO are sampled predominantly. Lower
             left panel: the free energy profile for the rotation
             between Fe--CO ($\theta = 0$) and Fe--OC ($\theta =
             180^\circ$) from the simulation with multipoles on the
             CO. Lower right panel: free energy profile along $\theta$
             with a point charge model for CO. Black trace is the raw
             data and red trace a running average.}
\label{fig:fig2}
\end{center}
\end{figure}

\noindent
The spectroscopy and reaction dynamics of photodissociated CO and NO
in Myoglobin (Mb) has been investigated intensely over the past $\sim
50$ years. For the present discussion, the infrared spectroscopy and
rebinding dynamics (see next section) is of primary
interest. Experimentally, the IR spectrum of the photodissociated CO
ligand was found to exhibit a split spectrum with two peaks separated
by $\sim 10$ cm$^{-1}$ which was associated with two conformational
substates. Despite some early efforts,\cite{meller:1998,anselmi:2007}
the assignment of these two states was unclear until simulations with
sufficiently detailed electrostatic models for the free ligand were
used.\cite{MM.mbco:2003,MM.mbco:2008} Together with subsequent
mutation studies\cite{nienhaus:2005} the more re-shifted peak was
associated with the Fe--OC orientation whereas the less red-shifted
peak corresponds to the Fe--CO state.\cite{MM.mbco:2006} Figure
\ref{fig:fig2} shows the overall IR spectrum of the free CO ligand
(red dashed line) together with spectra for the two suspected
conformational substates (Fe--CO in green; Fe--OC in blue). Firstly,
the total spectrum agrees favourable with the experimentally reported
spectrum. Secondly, the state-specific spectra suggest that although
they provide an identification of the two states, while sampling one
substate, e.g. Fe--CO, the spectroscopy is still sensitive to the
presence of the second state, i.e. Fe--OC. The reason for this is the
low isomerization barrier between the two states which is reported in
the lower left panel in Figure \ref{fig:fig2}. The Fe--CO state
($\theta = 0$) is somewhat more stable than the Fe--OC orientation
($\theta = 180^\circ$) and the two states are separated by a barrier
of $\sim 0.7$ kcal/mol, compared with the experimentally reported
barrier of 0.5 kcal/mol.\cite{kriegl:2003} The lower right panel in
Figure \ref{fig:fig2} shows the free energy profile along the CO
rotation angle when using a point charge model and demonstrates the
superiority of a more refined interaction model based on atomic
multipoles.\\

\section{Reaction Dynamics}
Following chemical reactions by way of computer simulations is a
challenging undertaking. Although in principle {\it ab initio} MD
(AIMD) simulations provide an obvious route, computational feasibility
and accuracy often preclude such an approach. Routine applications of
full AIMD simulations for reactions are typically limited to tens or
hundreds of trajectories covering 10s to 100s of picoseconds of
simulation time at the semiempirical or density functional theory
(DFT) level.\cite{MM.amm:2002,gerber:2006,gerber:2018}\\

\noindent
For tri- or tetra-atomic systems the investigation of reaction
dynamics at the classical or quantum nuclear dynamics level has an
extensive history, dating back at least to the earliest H+H$_2$
hydrogen exchange simulations\cite{karplus:1964} based on a reactive
London-Eyring-Polanyi-Sato (LEPS) surface.\cite{london29,lep31,leps2}
One surprising finding was that such quasi-classical trajectory (QCT)
simulations agree quite well with rigorous quantum
simulations\cite{schatz:1976} despite the fact that the H+H$_2$ system
is particularly susceptible to quantum effects including zero point
motion and tunneling.\\

\noindent
With the advent of efficient, high-level electronic structure
calculations the configurational space of small systems could be
covered adequately by computing energies for many geometries. As a
consequence, empirical PESs were largely superseded and the problem
shifted to representing the computed points such that the total
potential energy can be evaluated with comparable accuracy as the
underlying quantum chemical calculations. Fitting full-dimensional
PESs even for triatomic systems is still a formidable task and with
increasing number of dimensions becomes progressively more
difficult. One difficulty that is encountered and detrimental to the
dynamics of the system is the presence of ``holes'' in the
PES.\cite{fortenberry:2015,poirier:2019,poirier:2020,bowman:2021}
Holes may even develop in non-parametric, machine-learned energy
functions.\cite{tkatchenko:2020,bowman:2021} On the other hand
kernel-based representations with physically meaningful asymptotic
decay for large separations are less prone for developing irregular
features.\cite{ho96:2584,unk17:1923}\\

\noindent
Recent studies of gas-phase reaction dynamics have used
machine-learned PESs based on permutationally invariant polynomials
(PIPs),\cite{braams2009permutationally} PIPs using neural networks
(NNs),\cite{jiang2016potential} Gaussian Processes,\cite{guan:2018}
reproducing kernel Hilbert space (RKHS)
representations,\cite{ho96:2584,unk17:1923} or deep NNs such as
PhysNet or DeePMD.\cite{MM.physnet:2019,car:2018} A topical
application of a PIP-fitted PES to high-level electronic structure
calculations (MRCI, CCSD(T), and CCSD(T)-F12)\cite{wang:2017}
concerned the characterization of rotational resonances in the H$_2$CO
roaming reaction.\cite{quinn:2020} The particular interest in this
study was to provide an explanation for the origin of the
experimentally observed bimodal distribution of the CO fragments
following the decay of the excited H$_2$CO reactant according to
[H$_2$CO]*$\rightarrow$[H+HCO]*$\rightarrow$[H$_2$]*+[CO]*. In the
second step roaming of the H-atom eventually leads to abstraction of
the hydrogen from HCO to form the products. Although the QCT
simulations did not quantitatively reproduce the experimentally
measured rotational distributions $P(j)$, their bimodal structure was
correctly captured. Analysis of the trajectories revealed that the
low-$j$ portion of $P(j)$ passes through a cis-like OCH$\cdots$H
geometry whereas the high-$j$ trajectories rather sample a trans-like
OCH$\cdots$H geometry. This provides a detailed understanding for the
relationship between the topography of the PES, the dynamics supported
by it, and how this is reflected in the experimental observables.\\

\noindent
Another area where ML-based PESs have been extensively used is in the
field of high-energy reactions such as in combustion and in
hypersonics. The DeePMD NN-based PES was used to understand the
combustion of methane.\cite{zeng:2020} From a 1 ns long simulation
almost 800 reactions were recorded. In addition, the simulations
discovered previously unknown reaction pathways and products. For
example, cyclopropene was formed in a sequence of reactions involving
CH$_2$CO+CH$_3$ to form cycloprop-2-en-1-one which, after colliding
with another CH$_3$, formed CH$_3$CCH$_2$ and finally stabilized as
cyclopropene after further hydrogen loss. This also highlights that
atomistic simulations based on machine learned PESs can be used to
discover new reactions, provided that the statistical model covers
chemical space sufficiently well and provides realistic reaction
energetics.\\

\noindent
The investigation of hypersonic re-entry is a challenging endeavour on
multiple length and time scales. Hypersonic flight is characterized by
shock waves with temperatures up to 20000 K. The chemistry is
dominated by atom+diatom reactions with the reactant molecules often
in highly excited ro-vibrational states. Because the flow contains
species such as N, O, NO, N$_2$, O$_2$, CO, or CN, reaction networks
are required to describe the chemical evolution of the flow. This is
often based on thermal rates $k(T)$ which can, however, not be
determined experimentally for the entire temperature range
required. Instead, rates can be determined from high-level PESs and
corresponding QCT simulations.\\

\begin{figure}
 \begin{center}
 \resizebox{0.99\columnwidth}{!}
           {\includegraphics[scale=0.1,clip,angle=0]{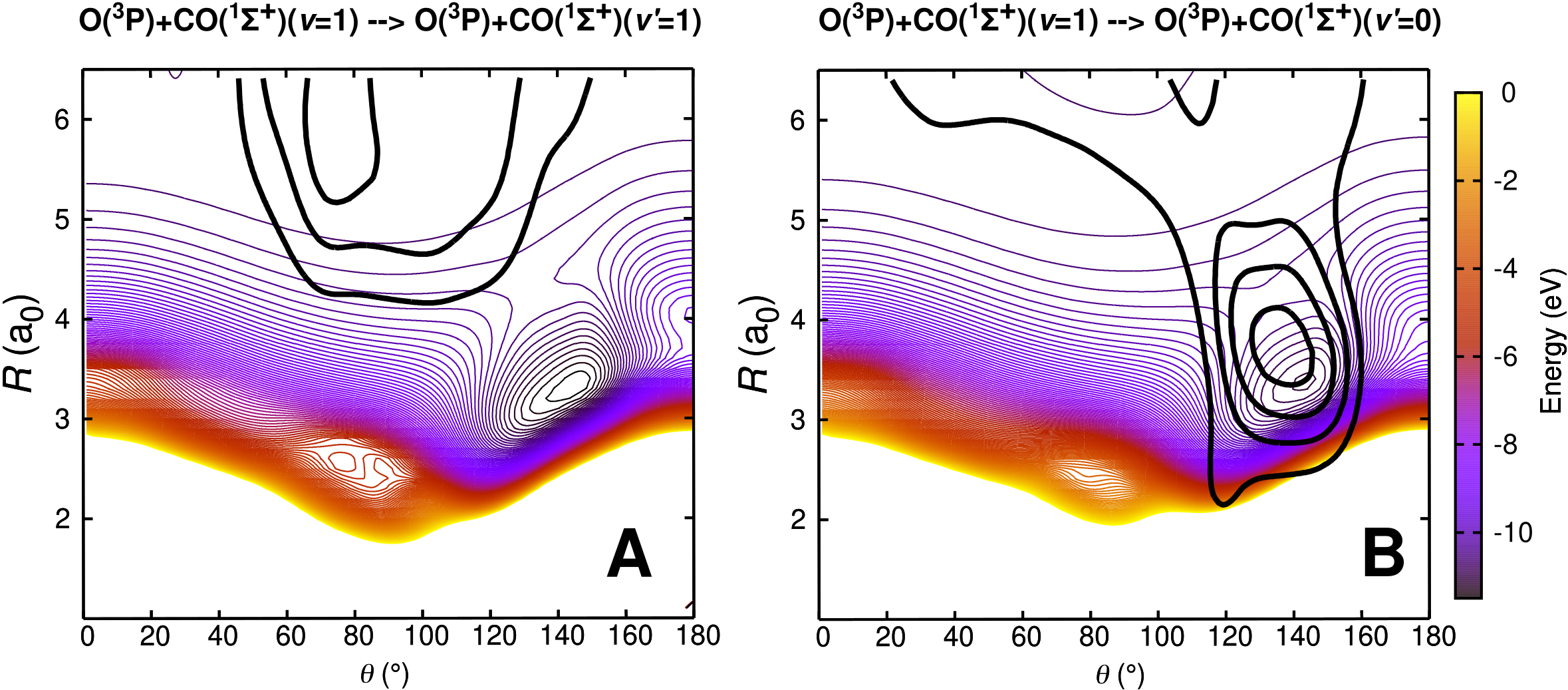}}
           \caption{Vibrational relaxation for the CO($^{1}\Sigma^{+};
             v=1$)+O($^{3}$P) collision. The underlying PES is for the
             $^3$A$'$ state with an equilibrium geometry for bent
             CO$_2$ with $\theta_{\rm OCO} \sim 140^\circ$. Panel A
             shows the density distribution from all non-relaxing
             (O($^{3}$P)+ CO($^{1}\Sigma^{+};
             v=1$)$\rightarrow$O($^{3}$P)+ CO($^{1}\Sigma^{+}; v'=1$))
             and panel B for all relaxing (O($^{3}$P)+
             CO($^{1}\Sigma^{+}; v=1$)$\rightarrow$O($^{3}$P)+
             CO($^{1}\Sigma^{+}; v'=0$)) trajectories.}
\label{fig:fig3}
\end{center}
\end{figure}

\noindent
This has been done for the [NOO],\cite{MM.no2:2020}
[NNO],\cite{MM.n2o:2020} [CNO],\cite{MM.cno:2018} and
[COO]\cite{MM.co2:2021} systems based on full-dimensional, reactive
RKHS PESs and QCT simulations. In all cases thermal reaction and
vibrational relaxation rates are in good agreement with existing
experiments for temperatures up to $T \sim 5000$ K which allows
extrapolation to higher temperatures. As an example of the additional
information atomistic simulations can provide the example of
non-reactive and reactive vibrational relaxation of CO in collisions
with atomic oxygen is considered. For the non-reactive case,
non-relaxing trajectories O+CO$(v=1)$ $\rightarrow$ O+CO$(v'=1)$
collide predominantly in a T-shaped geometry and do not enter the
strongly interacting region of the collision complex, see Figure
\ref{fig:fig3}A. This differs for vibrationally relaxing trajectories
O+CO$(v=1)$ $\rightarrow$ O+CO$(v'=0)$ for which all collide in a
nearly-linear geometry O$\cdots$CO and enter the strongly interaction
part of the PES in panel B.\cite{MM.co2:2021} Similar to the bimodal
product state distribution of CO following the decay of
H$_2$CO,\cite{quinn:2020} particular reaction channels can be
associated with sampling of specific parts of the PES.\\

\noindent
Finally, reactive PESs have also been used for applications in
atmospheric chemistry. One early example is the investigation of
vibrationally induced reactivity in sulfuric acid which has been
proposed based on experiments\cite{vaida:2003} and demonstrated in a
series of atomistic simulations using AIMD\cite{Miller:2006} or
reactive MD
techniques.\cite{MM.h2so4:2011,MM.armd:2014,reyes.pccp.2014.msarmd}
More recently, the Criegee intermediate and its decomposition to
CH$_2$COH+OH has raised considerable interest due to both, its
relevance in atmospheric processes, in particular in
ozonolysis,\cite{criegee1949ozonisierung,alam2011total,novelli2014direct,taatjes2017criegee,mauldin2012new}
and the availability of a wealth of experimental
data.\cite{fang:2016,lester:2016,green2017selective} A first study was
based on a PES fit to PIPs and used in quasiclassical trajectory (QCT)
calculations. This PES was fit to $\sim 157000$ electronic energies (a
mixture of CCSD(T) and CASPT2) and covered \textit{syn}-CH$_3$CHOO,
the transition state (TS) to vinyl hydroxyperoxide (VHP), several exit
channel wells and the OH+vinoxy products. The QCT calculations were
initiated at the TS separating the \textit{syn}-CH$_3$CHOO and VHP
wells instead of the syn minimum. This was done owing to the long
lifetime of the energized \textit{syn}-CH$_3$CHOO and the large
computational effort needed to propagate the trajectories.\\

\noindent
In a more recent exploration\cite{MM.criegee:2021} of the full
reaction pathway between \textit{syn}-CH$_3$CHOO and the CH$_2$COH+OH
product a MS-ARMD-based PES surface was fit to several thousand
reference energies and a PhysNet-PES was generated for $\sim 100000$
energies, both at the MP2 level of theory. Next, an aggregate of $\sim
10^5$ classical trajectory calculations from thermal ensembles at
different temperatures of the \textit{syn}-CH$_3$CHOO reactant state
were carried out to follow the reactive dynamics after vibrational
excitation of the CH stretch with two vibrational quanta. The energy
dependence of the experimentally measured rates for OH-elimination
were realistically described from such a computational approach.\\

\noindent
Reactive dynamics simulations can also been carried out in
solution. One example is the F + CD$_3$CN abstraction reaction to form
DF+CD$_2$CN in CD$_3$CN which was treated with a multi-state empirical
valence bond (EVB) approach.\cite{harvey:2015} This work reproduced
the experimentally observed\cite{orr:2015} solvatochromic shift of the
DF product in CD$_3$CN and reported that the DF product contains a
significant amount of vibrational energy ($v = 2-3$). EVB-based
approaches have also been employed extensively in biomolecular
simulations.\cite{aqvist:1993,kamerlin:2010,mavri:2014,kulkarni:2018,planas:2021}
From a ML perspective, RKHS-represented PESs have been used akin to
QM/MM simulations for ligand rebinding reactions of ligands to
proteins, in particular CO and NO to heme-containing
proteins.\cite{MM.cco:2014,MM.mbno:2015,MM.mbno:2016}\\

\section{Concluding Remarks and Prospects}
So far the focus has been on carrying out atomistic simulations with
input from machine learning at the level of the energy function, its
parametrization and its representation.\cite{unke:2021,MM.pes:2020}
While this aspect will gain in relevance in the time to come, {\it
  analysis} of MD-generated trajectory is an equally important
aspect. After all, the analysis of such trajectories provides the
information to generate observables that can be compared with
experiments. This comparison is the necessary step for validation of
the computations with the laboratory experiments or from
observations.\\

\noindent
An early procedure for extracting progression coordinates from MD was
based on a neural network and was applied to alanine
dipeptide.\cite{ma:2005} This work employed a genetic neural network
to extract a minimal set of internal degrees of freedom to describe
the C$_{\rm 7eq}$$\rightarrow$$\alpha_{\rm R}$ transition for the
solvated dipeptide. For this, the committor probability was used as the
target function to be optimized. For the transition in the gas phase
two descriptors (two dihedral angles) were sufficient whereas for the
transition in solution three coordinates were required (one dihedral,
one intramolecular separation and the torque around one C-N
bond). More recently, a generalization of this framework based on
artificial NNs was presented.\cite{jung:2019}\\

\noindent
Machine learning was also used to determine essential internal
coordinates from extended protein MD simulations.\cite{brandt:2018}
The decision tree-based method (XGBoost)\cite{xgboost} was used to
characterize metastable states of heat shock protein-35 and
transitions between them and to the open/closed transition in
Lysozyme. Because the ML-based approach directly operates on a pool of
candidate features (typically atom distances, valence angles, dihedral
angles) the analysis provides an importance score for each of the
features for a transition between two states. This differs from the
more widely used principle component analysis (PCA) which only yields
the linear combination of such variables. For Lysozyme analysis by
XGBoost clarified that the distance $d_{4,60}$ between residues Phe4
and Lys60 is involved in the open-to-closed transition and that the
more obvious separation between residues Thr21 and Thr142 is not a
suitable progression coordinate. Importantly, PCA did not find the
relevance of $d_{4,60}$ which suggests that ML-based methods can be
profitably used to uncover mechanistic aspects of functional protein
motions.\\

\noindent
For chemical reactions ML-based methods were recently used to study
determinants in the reaction well that do or do not lead to chemically
productive (i.e. reactive) trajectories.\cite{tidor:2019} This study
confirmed that results from machine learning confirm earlier proposals
concerning the importance of electrostatic preorganization, or
enzyme-stabilized ``near-attack'' conformations as relevant for
enzymatic activity. As MD simulations provide both, coordinate and
velocity trajectories it is of interest to query whether one or the
other is more suitable to predict relevant degrees of freedom. It was
found that both types of information can distinguish reactive from
almost-reactive trajectories and that their combination performs even
slightly better. It is of interest to juxtapose the insight that while
still sampling the reactant well it is possible to predict whether or
not a particular trajectory is likely to react with explicit sampling
of reactive vs. non-reactive initial conditions.\cite{MM.mdp:2019}
Analysis of the minimum dynamical path for reactions in the gas phase
showed that specific combinations of initial position and momenta in
the reactant state lead to reactive trajectories whereas initial
conditions outside this manifold do not react.\\

\noindent
Another application of ML methods to chemical reactions is the
prediction of reaction rates. This has been done directly for rates
from a library of $\sim 40$ bimolecular reactions for which
$T-$dependent rates from transition state theory (TST), the Eckart
correction to TST, and a set of tabulated ``accurate rates'' from
2-dimensional calculations at 8 temperatures is
available.\cite{houston:2019} This data was used to learn a correction
to the product of the TST-rate and the Eckart correction by using
Gaussian process regression. The results for reactions not used in the
learning procedure indicate that it is possible to obtain thermal
rates close to those from explicit quantum simulations (using
MCTDH)\cite{mctdh} or trajectory-based quantum calculations (ring
polymer MD).\cite{craig:2005}\\

\noindent
An alternative procedure was followed from determining state-to-state
(STS) reaction probabilities for bimolecular reactions and computing
the thermal rates from the final state distributions.\cite{MM.nn:2019}
From quasiclassical trajectory (QCT) simulations for the
N($^4$S)+NO($^2\Pi$) $\rightarrow$ O($^3$P)+N$_2$(X$^1\Sigma_g^+$)
reaction the state-to-state cross sections $\sigma_{v,j \rightarrow
  v'j'} (E_t)$ as a function of the translational energy $E_t$ were
explicitly determined for 1232 initial states which amounted to $\sim
10^8$ QCT trajectories in total compared with an estimated $10^{15}$
QCT trajectories required for brute-force sampling of the
problem. This information was used as input to train a NN together
with features such as the internal energy, the vibrational and
rotation energy of the diatoms, or the turning points of the
diatoms.\cite{MM.nn:2019} From the NN-predicted STS cross sections the
thermal rate is obtained from integrating over all $v-$ and $j-$
states and $E_t$ using Monte Carlo integration. Comparison with rates
directly determined from QCT simulations shows that the trained NN
reaches accuracies better than 99 \% over a wide temperature range
$(1000 \leq T \leq 20000)$ K. A very recent generalization was
concerned with learning entire product state distributions for
specific reactant states (state-to-distribution model;
STD).\cite{MM.nn:2021} The performance of this model is equally good
as the STS model but at considerably reduced computational cost. At
the most coarse-grained level, NNs can also be used to learn product
state distributions from distributions of initial reactant states from
which again thermal rates can be determined.\cite{MM.nn:2020}\\

\noindent
It is also possible to apply machine learning methods to
spectroscopy. One recent application is to learn spectroscopic maps
that are used for 1- and 2-dimensional infrared
spectroscopy.\cite{skinner:2019} Gaussian regression was used for a
$\Delta-$learning-based approach to improve an initial set of
transition frequencies and dipole derivatives for the water-OH
stretch. Compared with a conventional, parametrized spectroscopic map
the machine learned model reduced the errors for the transition
frequencies and the dipole derivatives by a factor of $\sim 2$. The
additional computational overhead of the ML-based model compared with
a conventional spectroscopic map is a factor of $\sim 10^4$.\\

\begin{figure}
 \begin{center}
 \resizebox{0.99\columnwidth}{!}
           {\includegraphics[scale=0.1,clip,angle=0]{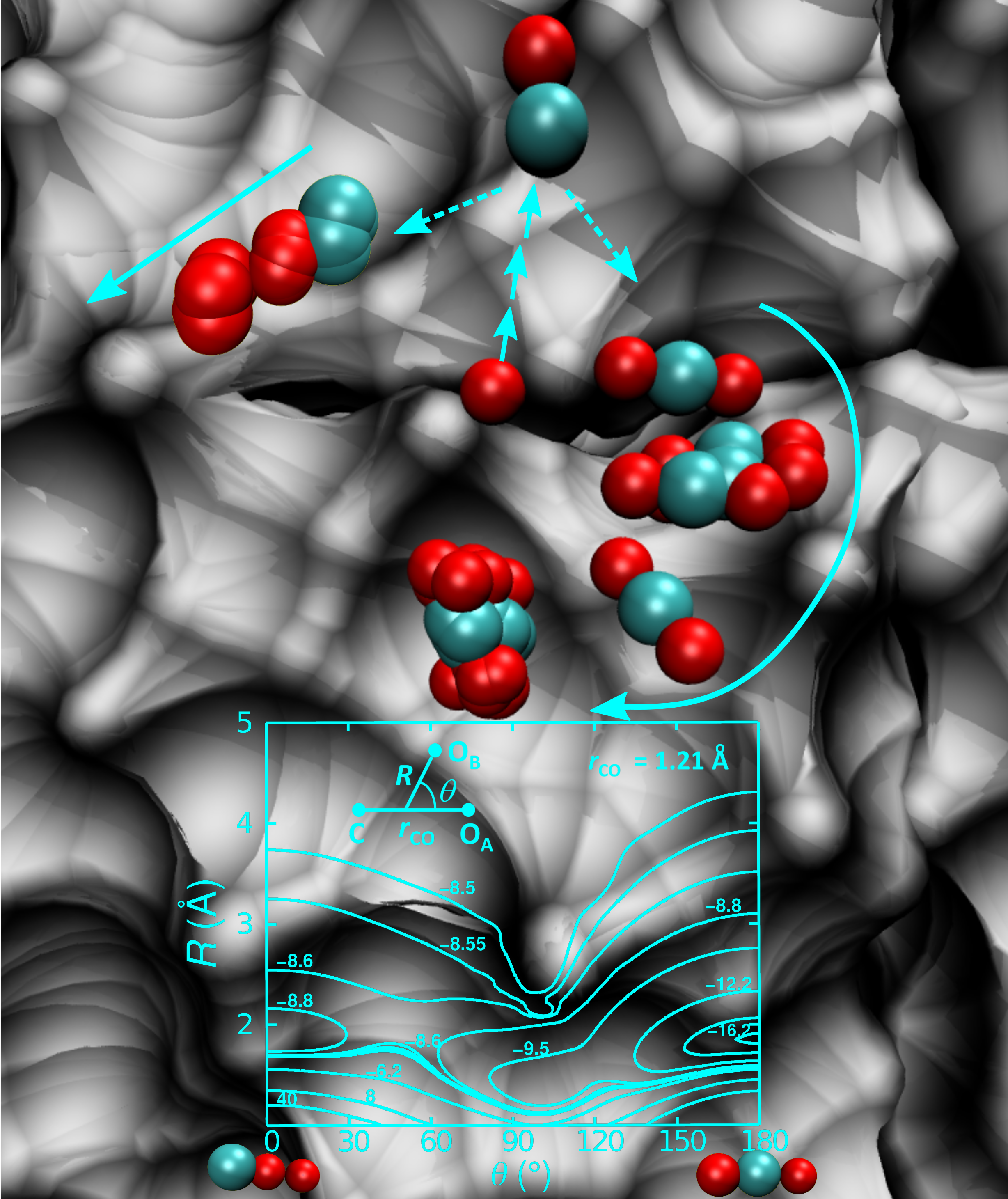}}
           \caption{CO+O recombination on amorphous solid water. The
             ground state PES supports both, the global OCO and the
             local COO minimum energy structure. Both are formed and
             stabilized from reactive MD simulations using the RKHS
             interpolated MRCI+Q/aug-cc-pVTZ reference energies.}
\label{fig:fig4}
\end{center}
\end{figure}

\noindent
Exploration of reaction mechanisms is another future application of
ML-based techniques. This has, for example, been noted when
investigating the combustion of methane.\cite{zeng:2020} Atmospheric
and interstellar chemistry are further fields for which this can
become relevant. For example, considering the O($^1$D)+CO($^1
\Sigma^+$) reaction on amorphous solid water
(ASW).\cite{MM.coo:2021,MM.coo:2022} Using a RKHS-based reactive PES
for the recombination reaction it was not only demonstrated that
CO$_2$ can form and stabilize on ASW. Consistent with earlier
experiments\cite{thrush:1973} the simulations also show that the COO
isomer can form and stabilize, see Figure \ref{fig:fig4}. Other recent
applications of ML-based energy functions to interstellar chemistry
concern the diffusion of atomic oxygen\cite{MM.odiff:2018} and
nitrogen\cite{kaestner:2020} and the formation of O$_2$ on
ASW.\cite{MM.o2:2019,MM.o2:2020}\\

\noindent
Atomistic simulations will continue to benefit from synergies with
machine learning technology. This includes both, the implementation
and technical advances of ML. From a computational perspective it is
desirable to fully integrate ML PESs into existing MD codes for
efficiency reasons and not to use a ML PES as an external energy
function akin to QM/MM calculations. From a technological
perspective, rigorous application of transfer- or $\Delta-$learning
approaches offer much scope to go beyond DFT-based PESs in an
efficient manner. First steps in this direction have been
undertaken.\cite{MM.tl:2021,MM.tl:2022}\\

\noindent
Another imminent improvement is the realistic description of long
range interactions. Most machine-learned PES are restricted to
exploration of the energetics around well-defined reference (minimum
energy) structures or for small reference systems using local
representations. However, for chemical reactions or systems in the
condensed phase the long range part of the intermolecular interaction
is essential and determining reference structures for input into
ML-based models becomes increasingly costly. This was recently
addressed within the DeepMD framework by using Wannier functions
centered on the atoms.\cite{e:2021} Alternatively, it may also be
possible to combine a short-ranged ML-PES with explicit long range
representations based on multipoles as has been done for ionic
complexes.\cite{MM.heh2:2019} On the other hand, with a suitable
high-level implementation machine learned PESs from reference
calculations in the gas phase may already be used in a QM/MM fashion
in MD simulations. Such ML/MM MD simulations provide considerable
speedup compared with conventional QM/MM simulations for sufficiently
high levels of the QM reference. Combined with transfer learning
techniques this will allow treatment of the solute at the correlated
level of theory (such as MP2 or even CCSD(T)). Here it is important to
stress that the computational cost for evaluating an ML PES is always
the same, irrespective of the level of theory it was trained on. The
computational cost for the different levels of theory used for the
reference calculations arises for the reference calculations which,
however, can be alleviated due to a) the massively parallel fashion
in which this step can be carried out and b) the fact that TL can
considerably reduce the number of high-level calculations that needs
to be carried out.\\

\noindent
In conclusion, with the ever increased computer power available,
atomistic simulations combined with tools from ML have a bright and
exciting future for exploring functional aspects of chemical
systems. This concerns in particular, but not exclusively, systems for
which state-of-the art experiments are available and can be carried
out. An MD/ML-based approach can also be used for prediction of
spectroscopic responses - e.g. upon protein mutation - or reaction
pathway and product exploration. This requires computationally
efficient implementations and reference calculations at sufficiently
high level of theory, both of which are becoming
available.\cite{MM.rkhs:2020,bowman.delta:2021,MM.tl:2022,bowman:2022}\\

\begin{acknowledgement}
This work was supported by the Swiss National Science Foundation
grants 200021-117810, 200020-188724, the NCCR MUST, and the University
of Basel.\\
\end{acknowledgement}

\section*{Author Biography}
{\bf Markus Meuwly} studied Physics at the University of Basel and
completed his PhD in Physical Chemistry working with
Prof. J. P. Maier. After postdocs with Prof. J. Hutson (Durham) and
Prof. M. Karplus (Strasbourg and Harvard) as a Swiss National Science
Foundation Postdoctoral Scholar he started as a F\"orderprofessor at
the University of Basel in 2002 where he is Full Professor of Physical
and Computational Chemistry. He also holds a visiting professorship at
Brown University, Providence, RI. His scientific interests range from
accurate intermolecular interactions based on multipolar, kernel- and
neural network-based representations to applications of quantitative
molecular simulations for cold (interstellar) and hot (hypersonics)
environments and the investigation of the spectroscopy and reactive
dynamics in proteins and in the condensed phase.\\

\bibliography{refs.tidy}
\end{document}